\documentclass[nofootinbib,aps,superscriptaddress,pra,10pt,twocolumn,longbibliography]{revtex4-2}

\usepackage{graphicx}
\usepackage{tikz}
\usepackage{tikz-cd}
\usetikzlibrary{decorations.markings,positioning}

\usepackage{braket}
\usepackage{float}  
\usepackage[english]{babel}
\usepackage[utf8]{inputenc}

\usepackage[colorlinks=true,linktocpage=true]{hyperref}
\hypersetup{
    allcolors  = {blue},
}
\usepackage{booktabs} 

\usepackage{amsmath}
\usepackage{amssymb} 
\usepackage{bbold}
\usepackage{amsthm}

\usepackage{xfrac}
\usepackage{physics}
\usepackage{centernot}

\usepackage{standalone}

\usepackage{listings}
\usepackage{footmisc}

\usepackage{microtype} 
\usepackage[capitalize]{cleveref} 

\usepackage{nomencl}
\makenomenclature

\usepackage{footmisc}

\usepackage{soul}

\usepackage{color}

\begin{document}

\title{Extended Wigner's friend paradoxes do not require nonlocal correlations}

\author{Laurens Walleghem}
\affiliation{Department of Mathematics, University of York, Heslington, York YO10 5DD, United Kingdom}
\author{Rafael Wagner}
\affiliation{International Iberian Nanotechnology Laboratory (INL), Av. Mestre Jos\'{e} Veiga, 4715-330 Braga, Portugal}
\affiliation{Centro de F\'{i}sica, Universidade do Minho, Braga 4710-057, Portugal}
\author{Y{\`i}l{\`e} Y{\=\i}ng}
\affiliation{Perimeter Institute for Theoretical Physics, Waterloo, Ontario, Canada, N2L 2Y5}
\affiliation{Department of Physics and Astronomy, University of Waterloo, Waterloo, Ontario, Canada, N2L 3G1}
\author{David Schmid}
\affiliation{International Centre for Theory of Quantum Technologies, University of Gda{\'n}sk, 80-309 Gda\'nsk, Poland}

\begin{abstract}
Extended Wigner's friend no-go theorems provide a modern lens for investigating the measurement problem, by making precise the challenges that arise when one attempts to model agents as dynamical quantum systems. Most such no-go theorems studied to date, such as the Frauchiger-Renner argument and the Local Friendliness argument, are explicitly constructed using quantum correlations that violate Bell inequalities. In this work, we show that such correlations are not necessary for having extended Wigner's friend paradoxes, by constructing a no-go theorem utilizing a proof of the failure of noncontextuality. The argument hinges on a novel metaphysical assumption (which we term Commutation Irrelevance) that is a natural extension of a key assumption going into the Frauchiger and Renner's no-go theorem.
\end{abstract}
\date{\today}

\maketitle


\section{Introduction}

In recent years, a number of extensions of the famous Wigner's friend thought experiment~\cite{Wigner1995} have been proposed~\cite{schmid2023review,brukner2017quantum,frauchiger2018quantum,bong2020strong}. These extended Wigner's friend (EWF) arguments aim to elucidate the difficulties that arise when one attempts to model an observing agent as a quantum system in their own right. Arguments of this sort were popularized in 2016 by Frauchiger and Renner~\cite{frauchiger2018quantum}, leading to debates on its assumptions, and on how different interpretations of quantum mechanics deal with such no-go theorems~\cite{nurgalieva2018inadequacy,sudbery2019hidden,vilasini2019multi,sudbery2017single,schmid2023review,relano2018decoherence,relano2020decoherence,kastner2020unitary,zukowski2021physics,gambini2019single,di2021stable,aaronson2018s,healey2018quantum,araujo2018flaw,drezet2018wigner,fortin2019wigner,losada2019frauchiger}.
These discussions provide a fresh angle on the measurement problem in quantum theory~\cite{baumann2016measurement,vilasini2022general,kastner2020unitary,brukner2018no,brukner2017quantum}.
A device-independent no-go theorem based on a similar EWF scenario has been developed, namely, the Local Friendliness no-go theorem~\cite{bong2020strong,haddara2022possibilistic,cavalcanti2021implications,wiseman2023thoughtful,ying2024relating}. A number of other no-go theorems have been introduced~\cite{leegwater2022greenberger,ormrod2022no,healey2018quantum,PuseyYoutube,ormrod2023theories}, although all of these are essentially variants of the Frauchiger-Renner or the Local Friendliness argument.   
For an introduction to Wigner's friend and various extensions, see Ref.~\cite{schmid2023review}.

As has been recognized numerous times~\cite{drezet2018wigner,aaronson2018s,montanhano2023contextuality,fortin2019wigner,vilasini2019multi,vilasini2022general,schmid2023review}, the Frauchiger-Renner construction is built around the correlations arising in Hardy's no-go theorem for Bell's notion of local causality~\cite{hardy1993nonlocality}. Indeed, the vast majority~\cite{bong2020strong,haddara2022possibilistic,frauchiger2018quantum,vilasini2019multi,vilasini2022general,ormrod2022no,leegwater2022greenberger,brukner2018no,utreras2022extended,wiseman2023thoughtful,brukner2017quantum} of EWF paradoxes are built around such nonlocal\footnote{ We use the standard term `nonlocal' here to describe correlations that violate Bell inequalities (despite the term's inherent bias~\cite{wolfe2020quantifyingbell}). } correlations--- that is, spacelike-separated correlations that cannot be explained by any classical common cause explanation~\cite{bell1964einstein,Bell_1976,Wood_2015,cavalcanti2014modifications,wolfe2020quantifyingbell,henson2014theory}. 

In this work, we construct a paradox that is close in spirit to Frauchiger and Renner's, but starting from contextual correlations for measurements on a single system---correlations which therefore do not involve Bell nonlocality in any way. To obtain such a paradox, it is necessary to extend one of the key assumptions made by Frauchiger and Renner slightly. However, we argue that the extended assumption has essentially the same motivations as Frauchiger and Renner's assumption, and so our no-go theorem has the same metaphysical consequences, despite not making use of nonlocal correlations.

Given the strong mathematical and conceptual connections between nonlocal and contextual correlations~\cite{Wright_2023,LIANG20111,Budroni_2022}, it is not surprising that EWF arguments involving contextual correlations would be possible (see also the subsequent work~\cite{walleghem2024connecting,walleghem2024strong}). But note that our construction here does not leverage these connections at all; indeed, it is based on contextual correlations that are \emph{not} nonlocal.  

There are two particular hurdles that need be overcome to construct a useful EWF argument based on contextual correlations (that are not also nonlocal). The first of these is that such an argument cannot assume noncontextuality itself, for if it were to do so, then it would merely be a worse version of a noncontextuality no-go theorem  (requiring strictly stronger assumptions than standard noncontextuality arguments). The second is that such an argument cannot apply the projection (collapse) postulate to any process which is treated unitarily at any point in the argument, as doing so significantly weakens any apparent contradiction (e.g., such arguments would not teach us anything new beyond Wigner's original thought experiment; see also \cite[Sec. III]{schmid2023review} and \cite{baumann2016measurement,sudbery2017single,lazarovici2019quantum,leegwater2022greenberger}). We expand on why these are hurdles, and how we overcome them, in the later sections of the manuscript.

There are a few related works that construct EWF arguments on a single system~\cite{allardGuerin2021nogotheorem,nurgalieva2023multi,szangolies2023quantum}. Refs.~\cite{nurgalieva2023multi,szangolies2023quantum} implement constructions similar to ours based on contextual correlations, but do not identify or motivate the critical assumption that (we will argue) is required.  In addition, Ref.~\cite{nurgalieva2023multi} applies the projection postulate to processes that are {\em also} treated unitarily in the argument, which (as mentioned just above) undermines the apparent paradox.  

Ref.~\cite{allardGuerin2021nogotheorem} also builds an EWF argument on a single system---one that is not based on contextual correlations, but rather on an assumption of a kind of linearity~\cite{allardGuerin2021nogotheorem}. This assumption can be criticized on interpretation-independent grounds~\cite{schmid2023review}.
Although the novel assumption required in our construction can also be criticized and would be rejected in some interpretations (as we discuss later), we argue that it may be motivated within others (and in particular, the same ones under which Frauchiger-Renner's assumptions are motivated). Still, understanding the problems with the linearity assumption in Ref.~\cite{allardGuerin2021nogotheorem} gives interesting insights---e.g., into the nature of quotiented representations of quantum processes~\cite{schmid2023review}, and so is worth study. Similarly, we hope that our no-go theorem will help provide insights into the metaphysical assumptions going into the Frauchiger-Renner argument and related arguments (with consequent implications for interpretations of quantum theory). 
In particular, if one grants that Frauchiger-Renner's (implicit~\cite{schmid2023review}) assumption of Timing Irrelevance is plausible, but wishes to reject our analogous assumption (termed Commutation Irrelevance, introduced below), then an explanation must be constructed for how the former and not the latter can be motivated within some given interpretation.

{
\section{Assumptions for EWF arguments}
}

A number of assumptions we consider here are common to most EWF arguments. One standard assumption is the \emph{Universality of Unitarity}, dictating that all dynamics can be described unitarily, even those of macroscopic systems including observers.  It follows that if an agent had sufficient (and extreme) technological capabilities, they could apply the inverse of the unitary describing a measurement to undo that measurement process. Such an agent is called a {\em superobserver}.
We also make the standard assumption of \textit{Absoluteness of Observed Events}~\cite{bong2020strong,haddara2022possibilistic,cavalcanti2021implications} that the outcome obtained in a measurement performed by an observer is single and absolute---not relative to anyone or anything.

Another standard assumption for EWF no-go theorems is that the outcome for any single measurement (or outcomes for measurements done in parallel, if those outcomes are jointly observable) occurs at frequencies obeying the Born rule, in accordance with operational quantum theory. Here we only need the \emph{Possibilistic Born Rule}, demanding that such outcomes never occur if the Born rule assigns probability 0 to them, and that otherwise, they sometimes occur.

We will make use of all the assumptions introduced above. Our no-go theorem also makes use of one novel assumption called \emph{Commutation Irrelevance}, which will be introduced in due time.\newline

\section{The $5$-cycle noncontextuality scenario}

We first summarize the 5-cycle noncontextuality no-go theorem from Ref.~\cite{cabello2013simple,klyachko2008simple,santos2021conditions}, on which our EWF argument will be based. 
Recall that a Kochen-Specker noncontextual assignment~\cite{Budroni_2022} for a set of measurements is a deterministic assignment of outcomes to those measurements, such that the assignment for each measurement is independent of which other compatible measurements are performed jointly with it. (The requirement of determinism can in turn be motivated by the more general assumption of generalized noncontextuality introduced by Spekkens~\cite{spekkens2005contextuality,spekkens2005contextuality}, which is defined and used later in this work.)

Consider five binary-outcome measurements $\{M_i\}_{i\in\{1,2,...,5\}}$, where the pairs $\{(1,2),(2,3),(3,4),$ $(4,5),(5, 1 )\}$ are jointly measurable. 
Imagine that the system is prepared such that the observations for these five joint measurements satisfy
\begin{subequations} \label{eq:5_cycle_12+23_34+45}
\begin{align}
 p(1,1|1,2)=0, \label{eq:5_cycle_12} \\
 p(0,0|2,3)=0, \label{eq:5_cycle_23} \\
    p(1,1|3,4) =0, \label{eq:5_cycle_34} \\
    p(0,0|4,5)=0, \label{eq:5_cycle_45}
    \end{align}
\end{subequations}
and
\begin{align}   \label{eq:01_box_not0}
            p(0,1|5,1) \neq 0.
    \end{align}

Denoting the deterministic assignment of the outcome of measurement $M_i$ as $m_i$, Eq.~\eqref{eq:5_cycle_12+23_34+45} implies that 
\begin{equation}
\label{eq:mcontr}
   m_1=1 \Rightarrow m_2=0 \Rightarrow m_3=1 \Rightarrow m_4=0 \Rightarrow m_5=1.
\end{equation}
Therefore, in every run of the experiment where the outcome of $M_1$ is $1$, the outcome of $M_5$ must be $0$. However, \cref{eq:01_box_not0} tells us that in some runs of the protocol, the outcome of $M_5$ is $0$ and the outcome of $M_1$ is $1$. Thus, there is no Kochen-Specker noncontextual assignment for these runs. 

The set of correlations in Eq.~\eqref{eq:5_cycle_12+23_34+45} and Eq.~\eqref{eq:01_box_not0} has a quantum realization as follows.
Consider a qutrit prepared in the state \begin{equation}\label{eq: eta}
    |\eta \rangle := \sqrt{\frac{1}{3}} (|0\rangle+\vert 1\rangle+\vert 2\rangle)\equiv \sqrt{\frac{1}{3}} (1,1,1)^T,
\end{equation} where $T$ denotes transposition, and define measurements 
\begin{equation}
\label{eq:M}
M_i:=\{\vert v_i\rangle \langle v_i \vert ,\mathbb{1}-\vert v_i\rangle \langle v_i \vert \},
\end{equation}
with the states $|v_i \rangle$ defined as \begin{equation}\label{eq: KCBS measurements}
    \begin{split}
        &|v_1 \rangle = \sqrt{\frac{1}{3}} (1,-1,1)^T, \\
        &|v_2 \rangle = \sqrt{\frac{1}{2}} (1,1,0)^T, \\
         &|v_3 \rangle = (0,0,1)^T, \\
        &|v_4 \rangle = (1,0,0)^T, \\
        &|v_5 \rangle = \sqrt{\frac{1}{2}} (0,1,1)^T.
    \end{split}
\end{equation} 
We take outcome $0$ of a given measurement to correspond to $\mathbb{1}-\vert v_i\rangle \langle v_i \vert $ and outcome $1$ to correspond to $\vert v_i \rangle \langle v_i \vert $. In this case, Eqs.~\eqref{eq:5_cycle_12}-\eqref{eq:5_cycle_45} are satisfied  (regardless of the state that the measurements act on) 
while $p(0,1|5,1) = \big|  \langle \eta | (\mathbb{1}-|v_5\rangle \langle v_5 |) |v_1 \rangle \big|^2 =  1/9\neq 0$, leading to Eq.~\eqref{eq:01_box_not0}. Hence, quantum theory is not consistent with noncontextuality.

\section{An EWF scenario with measurements on a single system}

We now leverage these 5-cycle correlations to construct an EWF no-go theorem. The argument involves five friends (denoted $A_1$, ..., $A_5$) who perform the five measurements $M_i$ defined in \cref{eq:M} sequentially on a single system $S$ prepared in the state $\ket{\eta}$ of \cref{eq: eta}, and a superobserver Wigner who undoes the first three of these measurements at particular times during the protocol.

By the assumption of Universality of Unitarity, each observer can be modeled as a quantum system---and indeed, as a qubit, since only two orthogonal (coarse-grained) states of the observer are relevant to the argument.
If each measurement $M_i$ is performed in a sufficiently isolated environment, Universality of Unitarity further implies that it can be described by a unitary. In particular, if the measurements are performed in a manner that has minimal disturbance on system $S$, then each measurement $M_i$ can be modeled by a CNOT gate $U_{M_i}$ between the system $S$ and the agent $A_i$, namely,{ 
\begin{align}
\label{eq_uni}
   &U_{M_i}:={\rm CNOT}_{SA_i} \\ \nonumber
   =&(\mathbb{1}-\ket{v_i}\bra{v_i})_{\rm S} \otimes\mathbb{1}_{\rm A_i}+   \ket{v_i}\bra{v_i}_{\rm S} \otimes ( \ket{1}\bra{0} + \ket{0}\bra{1})_{\rm A_i}.
\end{align}
}
where $\ket{0}_{A_i}$ is the coarse-grained state of $A_i$ having observed outcome 0 and $\ket{1}_{A_i}$ is for having observed outcome 1. We follow a standard abuse of notation by letting $\ket{0}_{A_i}$ also denote the coarse-grained `ready' state of the observer---that is, the initial state of each observer (prior to the measurement).

The superobserver Wigner is assumed to have perfect quantum control over the joint system of $A_1$, $A_2$, $A_3$, and $S$. In particular, we assume he has the (extreme) technological capabilities to implement the inverse of the unitary operations for the first three friends' measurements.

\begin{figure}[htb!]
\includegraphics[width=\columnwidth]{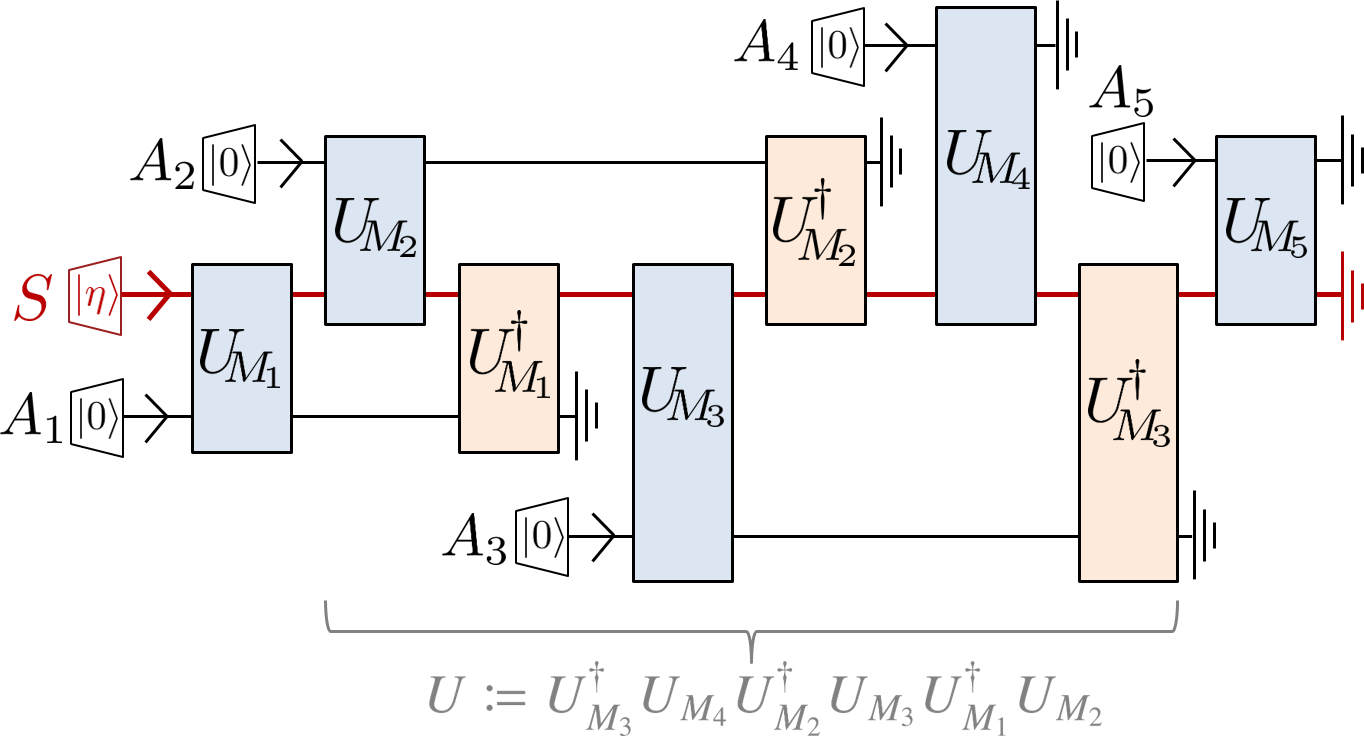}
    \caption{ \textbf{Schematic depiction of the scenario}. Five friends $A_1, \dots, A_5$ perform the five measurements from a 5-cycle noncontextuality no-go theorem on a system $S$, while a superobserver Wigner undoes the first three of these measurements at particular times in between.  Each measurement $M_i$ is modeled unitarily as $U_{M_i}$, following the assumption of Universality of Unitarity. The unitaries from $U_{M_2}$ to $U^{\dagger}_{M_3}$ are collectively denoted as $U$. In each measurement $M_i$, agent $A_i$ obtains an absolute outcome $a_i$, following the assumption of Absoluteness of Observed Events.}    \label{scenario}
\end{figure}
The exact sequence of operations is shown in the unitary circuit representation of the protocol in Fig.~\ref{scenario}.
Specifically, $M_1$ is the first measurement on $S$, followed by $M_2$, after which the superobserver undoes the first measurement by applying the unitary $U_{M_{1}}^{\dagger}$. Then $M_3$ is performed, followed by the undoing $U_{M_{2}}^{\dagger}$, followed by $M_4$, followed by the undoing $U_{M_{3}}^{\dagger}$, followed by $M_5$.

From Absoluteness of Observed Events, each agent $A_i$ observes an absolute outcome, denoted $a_i$, during their measurement $M_i$. We label the $\ket{v_i}\bra{v_i}$ outcome by $a_i=1$ and the other by $a_i=0$. 

\section{No-go theorem}

Measurements $M_1$ and $M_2$ are done immediately in sequence without any processes occurring in between, so the joint outcome $(a_1,a_2)$ can be directly observed (up until the time when $U_{M_2}^{\dagger}$ is performed).\footnote{ In fact, even though the records of the outcomes $a_1$ and $a_2$ may be erased later, it is nevertheless possible to obtain empirical evidence that the joint outcome $(a_1,a_2)$ agrees with the Born rule predictions. For example, we could add an extra instruction to friend $A_2$ in the experimental protocol: if she observes $a_1 = a_2 = 1$, in contradiction with \cref{eq_a12}, she should open the door of her lab and request that the experiment be halted immediately. Furthermore, in the case where $(a_1,a_2)$ is as expected, $A_2$ could write a note stating ``agrees with the Born rule'' and pass it to Wigner through a tiny slit in her lab door that is small enough and opened for a short enough time so that no other interaction between the inside and outside of the lab can occur. This note can be retained by Wigner even if the records of $(a_1,a_2)$ are later erased, and the rest of the experimental protocol can proceed unaffected. The inclusion of this note is analogous to the one in Deutsch's version of Wigner's friend~\cite[Sec. 8]{deutsch1985quantum}, the one in \cite[Sec.~IV.B]{copenhagen}, and the one in \cite[Sec. 4.2.2]{ying2024relating}.} 
 The Born rule predicts that 
\begin{equation}
\label{eq_a12}
    p(a_1=1,a_2=1)=0,
\end{equation}
just as in \cref{eq:5_cycle_12}, the correlation for the $(1,2)$ pair of measurements in the 5-cycle noncontextuality scenario. Moreover, this Born rule prediction---and those that follow---can be made without appealing to the projection postulate, as we explain later.

Measurements $M_2$ and $M_3$ are not done immediately in sequence; rather, the unitary $U_{M_1}^{\dagger}$ is applied in between them. Nevertheless, the joint outcome $(a_2,a_3)$ is still directly observable (up until $U_{M_2}^{\dagger}$ is performed). The Born rule predicts that
\begin{equation}
\label{eq_a23}
    p(a_2=0,a_3=0)=0,
\end{equation}
just as in \cref{eq:5_cycle_23}, the correlation for the $(2,3)$ pair of measurements in the 5-cycle noncontextuality scenario.
This is the case because the unitary $U_{M_1}^{\dagger}$ commutes with the $M_3$ measurement (as shown explicitly in Appendix~\ref{app_comUni}), and so it does not affect the Born rule prediction for $p(a_2,a_3)$. 

Similarly, the Born rule predicts that 
\begin{align}
    p(a_3=1,a_4=1)=0, \label{eq_a34} \\
    p(a_4=0,a_5=0)=0, \label{eq_a45}
\end{align}
where both refer to joint outcomes that can be directly observed, and the only process happening between each pair of measurements is a unitary that commutes with those measurements (and so cannot affect their statistics). \cref{eq_a34,eq_a45} are analogous to \cref{eq:5_cycle_34,eq:5_cycle_45}, respectively, in the 5-cycle noncontextuality scenario.

However, the same argument cannot be applied to the joint outcome $(a_1,a_5)$, since it necessarily cannot be observed by any observer, even in principle. This is because the outcome $a_1$ is erased (by the unitary $U_{M_1}^{\dagger}$) prior to the outcome $a_5$ being generated. Consequently, even though the process  $U:= U_{M_3}^{\dagger} U_{M_4} U_{M_2}^{\dagger} U_{M_3} U_{M_1}^{\dagger} U_{M_2}$, which is done between $M_1$ and $M_5$, commutes with $M_5$, the assertion that it does not affect the correlation between $M_1$ and $M_5$ is {\em not} guaranteed by the correctness of operational quantum theory (in particular, it does not follow from our assumption of Possibilistic Born Rule). Different interpretations of quantum theory may make different predictions for unobservable correlations, while still being consistent with operational quantum theory, as was highlighted already in, e.g., Ref.~\cite{schmid2023review}.

Therefore, to complete the argument, we require an assumption that the commutation relation $U_{M_5}UU_{M_1}=UU_{M_5}U_{M_1}$ implies $p_{U_{M_5}UU_{M_1}}(a_1,a_5)=p_{UU_{M_5}U_{M_1}}(a_1,a_5)$, where the former denotes the joint distribution when $U$ is performed between $M_1$ and $M_5$, and the latter denotes the joint distribution when $U$ is performed after $M_5$. Let us call this assumption {\em Commutation Irrelevance}. More generally, this assumption states that any unitary process performed between two measurements does not affect the correlations between their outcomes, provided that the unitary commutes with at least one of the two measurements---{\em even} if the outcomes of the two measurements are not jointly observable to any observer.\footnote{ We note that someone with operationalist leanings might object to this assumption on the grounds that it constrains correlations that are unobservable in principle. However, we note that it does not suffice to retreat to the position that `unperformed experiments have no results'~\cite{Peres1978}, since the relevant correlations in an EWF arguments refer to outcomes of {\em actually performed measurements}. This point was already made in e.g., Refs.~\cite{bong2020strong,cavalcanti2021implications,schmid2023review}.}

In the latter case, where $U$ is carried out {\em after} $M_5$, the correlation $p(a_1,a_5)$ is observable, and so is constrained by operational quantum theory to be
$p_{UU_{M_5}U_{M_1}}(a_1=1,a_5=0)=0$. Consequently,
assuming Commutation Irrelevance, the unobservable correlations for the actual experiment (where $U$ is done between $M_1$ and $M_5$) are  constrained in the same way:
\begin{equation}
    p(a_1=1,a_5=0)\neq0. \label{eq_a15}
\end{equation}
This is analogous to \cref{eq:01_box_not0} in the 5-cycle noncontextuality scenario.

\cref{eq_a12,eq_a23,eq_a34,eq_a45} imply that the actually observed outcomes must satisfy
\begin{equation}
    a_1=1\Rightarrow a_2=0 \Rightarrow a_3=1 \Rightarrow a_4=0 \Rightarrow a_5=1.
\end{equation} 
However, \cref{eq_a15} tells us that the event $(a_1=1,a_5=0)$ occurs in some runs of the protocol. 

Thus, we have a contradiction and consequently a no-go theorem against the conjunction of our assumptions: Universality of Unitarity, Absoluteness of Observed Events, Possibilistic Born Rule and Commutation Irrelevance.

In the appendix we extend these ideas to an infinite family of EWF paradoxes based on $n$-cycle proofs of the failure of Kochen-Specker noncontextuality for $n\geq 4$. We also introduce a unified description of possibilistic $n$-cycle contextuality models for both even and odd $n$.

{
\section{Discussion and outlook}
}

\subsection{How well motivated is Commutation Irrelevance?}

No-go results are only as interesting as their underlying assumptions are compelling.
Hence, it is necessary to question whether the assumption of Commutation Irrelevance is well-motivated. 
This is the key assumption beyond the standard assumptions common to EWF arguments, and plays a role analogous to the auxiliary assumptions of Timing Irrelevance or Local Agency in the Frauchiger-Renner and Local Friendliness arguments, respectively. 

In fact, the assumption of Commutation Irrelevance is a minor extension of the assumption of Timing Irrelevance~\cite{schmid2023review} required in the Frauchiger-Renner (and the Pusey-Masanes~\cite{PuseyYoutube,LeiferYoutube}) arguments. Timing Irrelevance is the assumption that the correlations obtained in two quantum circuits whose only difference is the timing of the measurements must be identical (even if the difference in timings matters for the question of whether or not the correlations in question are in principle observable). Timing Irrelevance is a special case of Commutation Irrelevance, since the identity channel (representing the action of waiting) commutes with any operation.

Both Timing Irrelevance and Commutation Irrelevance can be motivated by the assumption that quantum theory is {\em complete} in the sense that there is no deeper theory or deeper set of facts (like hidden variables) that determine what measurement outcomes occur. This assumption is endorsed by researchers sympathetic to the Copenhagen school of thought. On the other hand, as argued in Ref.~\cite{schmid2023review}, for interpretations that violate Completeness, or researchers who are less certain that quantum theory is the deepest possible description of nature, Timing Irrelevance and Commutation Irrelevance may not be compelling assumptions.

One explicit example of an interpretation that violates Commutation Irrelevance is Bohmian mechanics. This follows from the fact that Bohmian mechanics violates even the weaker assumption of Timing Irrelevance, as discussed in Ref.~\cite{schmid2023review}. One can also see Bohmian mechanics's violation of Commutation Irrelevance in the scenario discussed here by noticing that Bohmian mechanics reproduces quantum predictions for {\em observable} correlations. 
For our scenario (where $U$ is done between $M_1$ and $M_5$), Bohmian mechanics reproduces the observable predictions that $a_1=1\Rightarrow a_2=0 \Rightarrow a_3=1 \Rightarrow a_4=0 \Rightarrow a_5=1$.
Hence, $p_{U_{M_5}UU_{M_1}}(a_1=1,a_5=0)=0$ holds. However, for the case where $U$ is done after $M_5$, so that the outcomes of $M_1$ and $M_5$ can be jointly observed, Bohmian mechanics reproduces the quantum prediction for this correlation, namely, that $p_{UU_{M_5}U_{M_1}}(a_1=1,a_5=0)\neq 0$. So in Bohmian mechanics, $p_{U_{M_5}UU_{M_1}}(a_1=1,a_5=0)\neq p_{UU_{M_5}U_{M_1}}(a_1=1,a_5=0)$.

\subsection{Why one can (but should not) justify Commutation Irrelevance using noncontextuality}

A natural temptation is to justify Commutation Irrelevance by appealing to generalized noncontextuality~\cite{spekkens2005contextuality}. Specialized to quantum theory, generalized noncontextuality is the assumption that two operational processes that have the same quantum representation are modeled as identical stochastic processes in an ontological model. This assumption has been given much motivation in the literature~\cite{spekkens2005contextuality,schmidcharacterization2021,schmid2021guiding}, particularly by appealing to a version of Leibniz's principle~\cite{spekkens2019ontological} of the identity of indiscernible.  

To do so, consider again the process $U= U_{M_3}^{\dagger} U_{M_4} U_{M_2}^{\dagger} U_{M_3} U_{M_1}^{\dagger} U_{M_2}$ that is done between $M_1$ and $M_5$, and recall that  $U$ commutes with $M_5$.
Consequently, the measurement $M_5$ is operationally equivalent to the effective measurement obtained by first applying $U$ to one's system and then measuring $M_5$ (where in both cases we are considering $M_5$ as a terminal measurement).
In a noncontextual model, then, the response functions associated with these two measurements must be identical. As such, the correlations between $M_1$ and $M_5$ cannot depend on whether or not $U$ is done in between them. 

However, if one includes noncontextuality among one's assumptions in an EWF no-go theorem, then one's theorem will have no new foundational implications---rather, it will just become a worse proof of the failure of generalized noncontextuality. This is because {\em in addition} to the usual assumptions required in noncontextuality no-go theorems (chiefly, the assumption of tomographic completeness~\cite{spekkens2005contextuality,schmid2023addressing}), one also requires the baggage of superobservers.  Note that it is for entirely analogous reasons that the Local Friendliness no-go theorem~\cite{bong2020strong} assumes a weaker notion of locality than that used in Bell's theorem---if it used the same assumption Bell used, then it would teach us nothing new beyond Bell's theorem. 

As an aside, if one were content to assume noncontextuality, one could construct a no-go theorem that is considerably simpler---one which does not appeal to any commutation arguments, and where the only operational equivalences required are of the form $UU^{\dagger} = \mathbb{1}$.
Namely, one could merely consider an experiment wherein 
all five measurements ($M_1$ through $M_5$) are done in sequence, but with each measurement reversed by a superobserver in between. Then, because $M_i$ followed by $M_i^{\dagger}$ is operationally equivalent to the identity channel, generalized noncontextuality implies that the ontological representation of $M_i$ followed by $M_i^{\dagger}$ is given by the identity channel on the ontic state space. Consequently, all five measurements are performed on exactly the same ontic state, and so the fact that there is no consistent deterministic value assignment to these five measurements gives a contradiction.



\subsection{Why our argument does not appeal to the projection postulate}

The most obvious textbook quantum prescription for computing the predictions in \cref{eq_a12,eq_a23,eq_a34,eq_a45,eq_a15} makes use of the projection postulate (also known as the collapse postulate), since each such prediction involves two measurements implemented in sequence. However, we emphatically do {\em not} wish to appeal to the projection postulate. As argued in Ref.~\cite[Sec. III]{schmid2023review} and \cite{baumann2016measurement,sudbery2017single,lazarovici2019quantum,leegwater2022greenberger}, one should {\em expect} inconsistencies in any argument that requires one to treat a given process {\em both} unitarily {\em and} via the projection postulate, and so EWF arguments are only nontrivial if they ensure that for each given process in the argument, one gives {\em either} a unitary description {\em or} a projection postulate description. 

{This is related to a fairly common misconception about EWF arguments: it is often claimed that EWF arguments establish contradictions between unitary evolution and the collapse postulate. Indeed, early arguments (including Wigner's original thought experiment)~\cite{Wigner1995,deutsch1985quantum} did exactly this (by also implicitly making background assumptions essentially amounting to the eigenstate-eigenvalue link, see  \cite[Sec. III]{schmid2023review} for more details). However, it is in part for this reason that such arguments are easy to dismiss as unsurprising and easy to explain. One key part of what makes modern {\em extended} Wigner's friend arguments more compelling and less difficult to explain away is because they avoid applying two different models (unitary and nonunitary) to any single process in the argument. 

Rather, most modern EWF arguments signal an inconsistency between unitary evolution and {\em the Born rule}. Crucially, use of the Born rule does not imply any commitment to the projection/collapse postulate, nor indeed to any particular rule for updating states. The Born rule refers only to how one should compute predictions for the probabilities of outcomes, while state update rules (like the projection postulate, sometimes called the collapse postulate or the Luders-von-Neumann update rule) are a prescription for what quantum state one should assign to the system after the measurement. Assigning a state to a system after measurement is necessary if and only if one is interested in making predictions for {\em subsequent} measurements on the system.

Given that our protocol explicitly involves sequential measurements, it appears at first that we must appeal to some sort of state-update rule such as the projection postulate. However, we overcome this obstacle as follows.
}

Consider for example 
the correlation for $(a_1,a_2)$. We do not wish to compute this correlation via 
$p(a_1,a_2)=\Tr[\Pi_{a_2}(\Pi_{a_1}\rho\Pi_{a_1})]$ (where $\Pi_{a_i}$ denotes the respective projector) as this equation uses the projection postulate. 
Because measurement $M_1$ and $M_2$ are compatible, operational quantum theory gives a prescription for computing probabilities for their joint outcomes that does not involve the projection postulate. 
In particular, the probability distribution $p(a_1,a_2)$ for $M_1$ and $M_2$ measured in sequence (provided $M_1$ is measured in a minimally disturbing manner) can be inferred from the probabilities assigned to a single measurement in the eigenbasis which simultaneously diagonalizes $M_1$ and $M_2$; namely, the measurement $M=\{\ket{v_1}\bra{v_1},\ket{v_2}\bra{v_2},\mathbb{1}-\ket{v_1}\bra{v_1}-\ket{v_2}\bra{v_2}\}$. If we imagine a hypothetical experiment where Alice performs this three-outcome measurement, then one can compute the probabilities $p(a_1,a_2)$ for the actually performed measurements $M_1$ and $M_2$ from it. 

To be sure, the actual experiment does not involve the implementation of this joint measurement. We are only claiming that the statistics for the joint outcome $(a_1,a_2)$ can be computed by considering this hypothetical experiment, and that this computation does not require the projection postulate.

An analogous argument can be made for the correlation $p(a_2,a_3)$, but one must additionally note that the undoing operation $U_{M_1^\dagger}$ that is done between $M_2$ and $M_3$ cannot affect their outcome statistics, since it commutes with them, as noted earlier. A similar argument applies to $(a_3,a_4)$ and $(a_4,a_5)$.

For Eq.~\eqref{eq_a15}, the correlation $p(a_1,a_5)$ is unobservable to anyone, and so it is not constrained by operational quantum theory alone. But given Commutation Irrelevance, this unobservable correlation is equal to the observable correlation $p_{U_{M_1}U_{M_5}U}(a_1,a_5)$ in an experiment where $U$ is carried out after $M_5$ rather than before it. Because the latter correlation is observable, it is constrained by the predictions of operational quantum theory, and so we can again appeal to a measurement that jointly simulates $M_1$ and $M_5$ to compute the Eq.~\eqref{eq_a15} without appealing to the projection postulate.

This establishes how one can obtain Eqs.~\eqref{eq_a12}-\eqref{eq_a15} without using the projection postulate.

\subsection{Future directions.}

Exactly like the Frauchiger-Renner argument, our EWF no-go theorem is not an experimentally testable argument since not all correlations needed in the argument are observable. Nevertheless, it is a no-go theorem for quantum theory. It remains to be seen whether a device-independent argument of this sort can be constructed. Our results also open the possibility of new EWF scenarios constructed around other Kochen-Specker contextuality proofs. Our work also calls for further investigation into the importance of different forms of nonclassicality in EWF paradoxes.
{Such no-go theorems also motivate further study of the different assumptions that they show to be incompatible, and of how one can relax these assumptions to evade the no-go theorem (be it relaxing Absoluteness of Observed Events, Commutation Irrelevance, or background assumptions such as the fundamental and technological possibility of superobservers).}

\begin{acknowledgements}
We thank Ernesto Galvão, Rui Soares Barbosa, Leonardo Santos, Sidiney Montanhano, Vilasini Venkatesh, Nuriya Nurgalieva, Howard Wiseman, and Matthew Leifer for their comments and useful discussions.  LW also thanks Matt Pusey and Stefan Weigert for many discussions on the Frauchiger-Renner paradox.

RW acknowledges support from FCT – Fundação para a
Ciência e a Tecnologia (Portugal) through PhD Grant
SFRH/BD/151199/2021. LW acknowledges support from the United Kingdom Engineering and Physical Sciences Research Council (EPSRC) DTP Studentship (grant number EP/W524657/1). LW also thanks the International Iberian Nanotechnology Laboratory -- INL in Braga, Portugal and the Quantum and Linear-Optical Computation (QLOC) group for the kind hospitality. DS was supported by the Foundation for Polish Science (IRAP project, ICTQT, contract no. MAB/2018/5, co-financed by EU within Smart Growth Operational Programme). YY was supported by Perimeter Institute for Theoretical Physics. Research at Perimeter Institute is supported in part by the Government of Canada through the Department of Innovation, Science and Economic Development and by the Province of Ontario through the Ministry of Colleges and Universities. YY was also supported by the Natural Sciences and Engineering Research Council of Canada (Grant No. RGPIN-2017-04383).

\end{acknowledgements}

\begin{appendix}

\section{Kochen-Specker contextuality} \label{sec:KS_and_examples}

In this {appendix}, we introduce a unified presentation of $n$-cycle arguments for $n$ even and odd, we prove the commutation relations mentioned in the main text, and most importantly, we show that similar extended  Wigner's friend (EWF) paradoxes to those constructed in the main text exist, if instead of considering $A_1, \dots, A_5$ once considers $A_, \dots, A_n$, for any $n\geq 4$. 

\subsection{General framework} \label{sec:KS_contextuality}
Kochen-Specker contextuality is a specific property of probabilistic data defined over measurement scenarios. We define a \textit{measurement scenario}, also known as a \textit{compatibility scenario}, as any triplet $(X, \mathcal{M}, O)$ of a  set of ideal measurements $X$, a set of maximal contexts $C \in \mathcal{M}$~\footnote{A context $C \in \mathcal{M}$ is maximal if there exists no other set of compatible measurements $C'$ such that $C \subseteq C'$.}, with each $C$ corresponding to a set of compatible measurements (i.e., jointly measurable elements from $X$), and a set of labels for possible outcomes $O$ that each measurement can return. We will consider only finite sets of outcomes $O$, but scenarios with infinite outcomes can be properly studied~\cite{barbosa2022continuous}.

\begin{figure}
    \centering
    
    \includegraphics[width=\columnwidth]{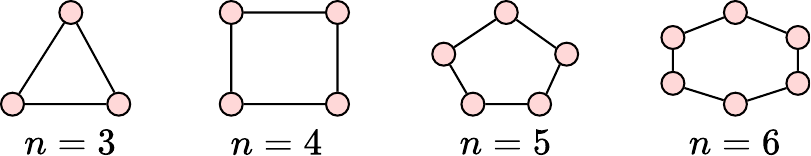}
    \caption{\textbf{$n$-cycle compatibility hypergraphs.} Cycle graphs $C_n$ for $n=3,4,5,6$. Each graph defines a compatibility structure that can be associated with measurement scenarios. Graph $C_4$ depicts the compatibility structure of the bipartite Bell scenarios leading to the Clauser-Horne-Shimony-Holt (CHSH) inequality~\cite{clauser1969proposed}. Graph $C_5$ depicts the compatibility structure leading to the Klyachko-Can-Binicioğlu-Shumovsky (KCBS) inequality~\cite{klyachko2008simple}.}
    \label{fig: cycles}
\end{figure}

In the compatibility hypergraph approach~\cite[Chapter 2]{amaral2018graph} (see also Ref.~\cite{amaral2014exclusivity}) we encode the information of maximal contexts $C \in \mathcal{M}$ of a given scenario $(X,\mathcal{M},O)$ in a hypergraph where each node represents an element of $X$ and hyperedges correspond to elements from $\mathcal{M}$, i.e., maximal contexts. In such a graph, hence, edges connect compatible measurements. In Fig.~\ref{fig: cycles} we depict the most studied and simplest non-trivial family of compatibility hypergraphs, known as $n$-cycle compatibility scenarios. Since in this case all hyperedges have only two elements, representing binary contexts, the hypergraphs reduce to graphs. Each such graph defines an infinite family of possible measurement scenarios, depending on the outcome set $O$ of the corresponding measurements.

A \textit{behavior} over a given finite measurement scenario is defined as a set of probability distributions over the maximal contexts, 
\begin{equation}\label{eq: behavior}
    \left\{p(\cdot |C): O^C\to [0,1]\bigr| \sum_{s \in O^C}p(s|C)=1, C \in \mathcal{M}\right\}.
\end{equation}
Each $s \in O^C$ is simply a tuple of possible outcomes resulting from jointly measuring a given system with all the ideal measurements in $C \in \mathcal{M}$. A behavior is said to be \textit{non-disturbing} whenever the marginals over intersections of contexts are equal, i.e., for all $C,C' \in \mathcal{M}$ such that $C\cap C' \neq \emptyset$ then $    \sum_{s \in O^{C};s|_{C \cap C'}=t}p(s|C) = \sum_{s' \in O^{C'};s'|_{C \cap C'}=t}p(s'|C')
$
for all $t \in O^{C \cap C'}$. The condition above becomes mathematically equivalent to the \textit{non-signalling} condition in case the compatibility structure of the scenario is the one from a Bell scenario. If there exists a single probability distribution $g$ over $X$ that correctly reproduces the behavior by marginals over contexts, i.e.,
\begin{equation}
    p(s|C) = \sum_{t \in O^X; t|_{C}=s}g(t)
\end{equation}
for all $C \in \mathcal{M}$ and $s \in O^C$, $g$ is said to be a \textit{global distribution} and the behavior $\{p(s|C)\}_{s \in O^C, C \in \mathcal{M}}$ is said to be \textit{noncontextual}. Hence, a behavior is said to be \textit{contextual} whenever such a global distribution $g$ cannot exist.  

The conditions above are theory-independent, i.e. they do not assume that a given behavior can be reached using quantum theory. We say that a behavior $\{p(s|C)\}$ defined with respect to a scenario $(X,\mathcal{M},O)$ has a \textit{quantum realization} whenever there exists some Hilbert space $\mathcal{H}$ for which each measurement in $M \in X$ corresponds to a projective measurement $\{P^M_{i}\}_{i=1}^{|O|}$, and such that there exists some density matrix $\rho$, i.e. a prepared state of some system, such that, letting $s = (o_1,\dots,o_n)$ and $C = \{M_1,\dots,M_n\}$ we have   $p(s|C) = \text{Tr}(\prod_{i=1}^n P_{o_i}^{M_i} \rho)$.

Let $\{p(s|C)\}$ be any behavior for a given scenario $(X,\mathcal{M},O)$. We define the \textit{possibilistic collapse} of this behavior as the set $\{\overline{p}(s|C)\}$ defined by the functions $\overline{p}( - | C ): O^C \to \{0,1\}$ for which 
\begin{equation}
    \overline{p}(s|C):= \left\{ \begin{matrix}
        1 & \text{if }p(s|C)>0\\
        0 & \text{if }p(s|C)=0
    \end{matrix} \right..
\end{equation}
We call $\{\overline{p}(s|C)\}$ a \textit{possibilistic behavior}~\cite{abramsky2011sheaf}. We say that a behavior $\{p(s|C)\}$ from a scenario $(X,\mathcal{M},O)$ is \textit{logically contextual} whenever there exists some $s \in O^C$ and some $C \in \mathcal{M}$ such that (i) $p(s|C)>0$, (ii) for $t \in \{u \in O^X : u|_C=s\}$ there exists some $C' \in \mathcal{M}$ such that $p(t|_{C'}|C') = 0$. In words, if there exists some possible joint outcome $s \in O^C$ for which all global assignments $t$ consistent with $s$ over $C$, i.e. satisfying $t|_C=s$, are impossible when restricted to some other context $C'$. Logically contextual behaviors lead to the so-called \textit{possibilistic paradoxes} as  merely the possibility of outcome results lead to paradoxical reasoning when assuming the existence of pre-determined values for all measurements simultaneously. Refs.~\cite{santos2021conditions,cabello2013simple} constructed possibilistic paradoxes for any $n$-cycle scenario where all measurements are dichotomic. 

\subsection{An $n$-cycle logical contextual model} \label{sec:n_cycle_logical_contextuality}
We present now the generalisation of the above 5-cycle logical contextual model to $n$-cycle logical contextuality scenarios, as shown in Refs.~\cite{cabello2013simple,santos2021conditions}. We provide a similar theory-independent operational description of an experiment consisting of $n$ boxes, which can be either full or empty, as before. Checking whether a box is empty or full is again considered as a measurement. The contexts, in this case the boxes that can be opened together, are the measurements of box $m \bmod n$ and $(m+1) \bmod n$, with the same compatibility structure from Fig.~\ref{fig: cycles}.  The  logically contextual $n$-cycle behaviours for $n$ odd or even are usually discussed in slightly different terms, as we review in \Cref{appendix:n_cycle_literature}. In \Cref{appendix: from odd to even} we show that how to transform the arguments in literature for odd and even $n$ into one single form, that is the one we present next.

The logically contextual behaviour for $n$-cycle scenarios, for any $n\geq 4$, is given by
\begin{align} 
&p(0,1 |i, i+1)=0 \text{ for $1 \leq i < n$}, \label{eq:cycle_for_all_n}
\end{align} and 
\begin{equation} \label{eq:cycle_n_neq0}
p(0,1 | 1,n) > 0.
\end{equation} These behaviours can be realized using quantum theory, from the behaviours of Ref.~\cite{santos2021conditions}, which have a quantum realisation as described in Ref.~\cite[Eqs. (49)-(55)]{santos2021conditions}, once some relabeling of outcomes is introduced, as we describe in detail in \Cref{appendix: from odd to even}. There, we also describe how the 5-cycle behaviour as described above in can be put into this form, when relabeling measurement outcomes.

\section{Logically contextual $n$-cycle behaviour for odd and even $n$} \label{appendix:n_cycle_literature}
In this appendix we start by presenting the logically contextual $n$-cycle behaviours from Ref.~\cite{santos2021conditions}. As we will see, the quantum realizations for  behaviours constructed for $n$ odd or even are slightly different. We will then show in Appendix~\ref{appendix: from odd to even} that a unified description for these specific behaviours exists, relaxing such apparent asymmetry between behaviors in even and odd scenarios. 
This unified description is the one we have used in the main text. 

\textit{An odd $n$-cycle logically contextual model.}-- We start presenting $n$-cycle logically contextual behaviors, with $n \geq 5$ an odd natural number. The system is prepared such that
\begin{align} 
&p(1,1 | 1,2)=0,\nonumber \\
&p(0,0| 2,3) = 0, \nonumber\\
&\hspace{1cm}\vdots \label{eq:odd_n_cycle_12+23_and_so_on}\\
&p(1,1 | n-2, n-1)=0, \nonumber\\
&p(0,0 | n-1, n)=0,\nonumber
\end{align}
and \begin{equation} \label{eq:odd_n_cycle_prob01_n1_nonzero}
p(0,1 | n, 1) > 0.
\end{equation} If one assumes that the result of finding a box empty or full is predetermined and independent of which boxes are opened (i.e. noncontextuality), then the first equations imply that \begin{equation} \label{eq:01_n1_zero}
    p(0,1|n,1) = 0,
\end{equation} in contradiction with Eq.~\eqref{eq:odd_n_cycle_prob01_n1_nonzero}. 
This model is thus logically contextual, with respect to the definitions provided in the main text, and has a quantum representation for any odd $n \geq 5$~\cite{cabello2013simple,santos2021conditions}.

\textit{An even $n$-cycle logically contextual model.}--We now construct a logically contextual $n$-cycle behavior, with $n \geq 4$ being an even natural number, following closely Ref.~\cite{santos2021conditions}. The behavior is defined such that 
\begin{align} 
&p(1,0|1,2)  =0, \nonumber\\
& \hspace{1cm}\vdots \nonumber\\
&p\left(1,0\Bigr|\frac{n}{2}-1,\frac{n}{2}\right)  =0, \nonumber\\
&p\left(1,1\Bigr|\frac{n}{2},\frac{n}{2}+1\right)   =0, \label{eq:even_n_cycle_12+23_and_so_on}\\
&p\left(0,1\Bigr|\frac{n}{2}+1,\frac{n}{2}+2\right) =0, \nonumber\\
& \hspace{1cm}\vdots \nonumber\\
&p(0,1|n-1,n)  =0,\nonumber
\end{align}
 and 
\begin{equation} \label{eq:even_n_cycle_prob_11_12_nonzero}
     p(1,1|n,1) > 0. 
\end{equation} If one assumes that the result of finding a box empty or full is predetermined and independent of which boxes are opened (i.e. noncontextuality), then Eqs.~\eqref{eq:even_n_cycle_12+23_and_so_on} imply that \begin{equation} \label{eq:even_01_n1_zero}
    p(1,1|n,1) = 0,
\end{equation} in contradiction with Eq.~\eqref{eq:even_n_cycle_prob_11_12_nonzero}. 
This behavior is logically contextual and has a quantum realization, for any even $n \geq 4$~\cite{santos2021conditions}. To do so, one can prepare a two-qubit system that satisfies Eqs.~\eqref{eq:even_n_cycle_12+23_and_so_on} and Eq.~\eqref{eq:even_n_cycle_prob_11_12_nonzero} when choosing appropriate projective measurements~\cite{santos2021conditions}.

\section{Turning an odd $n$-cycle paradox into an even $n$-cycle paradox and vice versa}\label{appendix: from odd to even}

In this appendix we show that by relabeling the outcomes $0 \leftrightarrow 1$ of measurements for $n$-cycle scenarios, we can transform the argument for odd $n$-cycle logical contextuality, as described in \Cref{appendix:n_cycle_literature}, in the same form as the argument for \textit{even} $n$-cycle logical contextuality from \Cref{appendix:n_cycle_literature}, and vice versa. In this way, we can present the logically contextual behaviours for odd and even $n$-cycles as presented in \Cref{appendix:n_cycle_literature} of \cite{santos2021conditions}, which are quantum realisable, in a uniform way.

\subsection{Relabeling outcomes in odd $n$-cycle logical contextuality}
 
Let $n$ be odd, and let us start from the odd $n$-cycle logical contextuality of Eqs.~\eqref{eq:odd_n_cycle_12+23_and_so_on}, \eqref{eq:odd_n_cycle_prob01_n1_nonzero}: 
 \begin{align} 
&p(1,1 | 1,2)=0,\nonumber \\
&p(0,0| 2,3) = 0, \nonumber\\
&\hspace{1cm}\vdots \label{eq:odd_n_start_appendix}\\
&p(1,1 | n-2, n-1)=0, \nonumber\\
&p(0,0 | n-1, n)=0,\nonumber
\end{align}
and \begin{equation} \label{eq:odd_n_start_appendix_neq_0}
p(0,1 | n, 1) > 0.
\end{equation}
Here, as before, $p(\cdot,\cdot | i,j)$ denotes the empirical probability distribution corresponding to the context where measurement $i$ and $j$ are jointly measured. 

Let us consider the following relabeling of measurements (recall that $n$ is odd): \begin{equation}
     \begin{split}
         M_1&: \text{relabel},\\
         M_2&: \text{keep}, \\
          M_3&: \text{relabel},\\
           M_4&: \text{keep},\\
            \vdots \\
        M_{n}&: \text{relabel},        
     \end{split}
 \end{equation} i.e., relabeling the outcomes of all odd measurements. By `relabel' of a measurement $i$, we mean that we are \textit{switching} the $0$ and $1$ labels for the outcomes of that measurement $i$, hence making $0 \leftrightarrow 1$. By `keep', we mean that no relabeling happens.
 The behaviour we end up with is the following:  
 \begin{align} 
&p(0,1 | i,i+1)=0 \text{ for all $1 \leq i < n$}, \label{eq:appendix_p_01_=0_forall_i} \\
&p(0,1| 1,n) > 0. \label{eq:appendix_p_01__neqzero} 
\end{align}
This model is logically contextual as, assuming non-disturbance (as we do throughout this paper), and the possibility to assign outcome values to all measurements simultaneously Eq.~\eqref{eq:appendix_p_01_=0_forall_i} implies that $p(0,1 \vert n,1)=0$ contradicting Eq.~\eqref{eq:appendix_p_01__neqzero}.

\vspace{0.5cm}

\textit{Example: relabeling outcomes in the 5-cycle logically contextual model.}-- As an illustration of the above relabeling method for odd $n$, we consider the logically contextual 5-cycle behavior as presented in the main text \cite{cabello2013simple}. We thus start from the behavior 
\begin{align}
 &p(1,1|1,2)=0,  \nonumber\\
 &p(0,0|2,3)=0, \nonumber \\
    &p(1,1|3,4) =0,  \label{eq:appendix_5_cycle} \\
    &p(0,0|4,5)=0, \nonumber
    \end{align}
and
\begin{align}  \label{eq:appendix_5_cycle_neq0}
            p(0,1|5,1) > 0.
    \end{align}
The relabeling we perform as explained in the previous section (where keep means do not relabel the outcomes of that particular measurement) is given by
 \begin{equation}
     \begin{split}
         M_1&: \text{relabel},\\
         M_2&: \text{keep}, \\
          M_3&: \text{relabel},\\
           M_4&: \text{keep},\\
        M_{5}&: \text{relabel},        
     \end{split}
 \end{equation} such that Eq.~\eqref{eq:appendix_5_cycle} becomes
  
\begin{align}
 p(0,1|1,2)=0,  \nonumber\\
 p(0,1|2,3)=0, \nonumber \\
    p(0,1|3,4) =0,  \\
    p(0,1|4,5)=0, \nonumber
    \end{align}
and Eq.~\eqref{eq:appendix_5_cycle_neq0}
\begin{align}  
            p(0,1|1,5) > 0.
    \end{align} We see that this indeed leads to the desired result.

\subsection{Relabeling outcomes in even $n$-cycle logical contextuality}
For even $n$, we can also reformulate the even $n$-cycle logically contextual behavior of \Cref{sec:n_cycle_logical_contextuality} into an argument similar to the behaviour obtained from odd $n$-cycle scenario as in the previous section by relabelling some outcomes. More precisely, let $n$ be even and let us start from the behavior as in Eqs.~\eqref{eq:even_n_cycle_12+23_and_so_on}, \eqref{eq:even_n_cycle_prob_11_12_nonzero}: \begin{align} 
&p(1,0|1,2)  =0, \nonumber\\
& \hspace{1cm}\vdots \nonumber\\
&p\left(1,0\Bigr|\frac{n}{2}-1,\frac{n}{2}\right)  =0, \nonumber\\
&p\left(1,1\Bigr|\frac{n}{2},\frac{n}{2}+1\right)   =0, \label{eq:even_n_start_appendix}\\
&p\left(0,1\Bigr|\frac{n}{2}+1,\frac{n}{2}+2\right) =0, \nonumber\\
& \hspace{1cm}\vdots \nonumber\\
&p(0,1|n-1,n)  =0,\nonumber
\end{align}
 and 
\begin{equation} \label{eq:even_n_start_appendix_neq0}
     p(1,1|n,1) \neq 0. 
\end{equation}

Let us consider the following relabeling of measurements (recall that $n$ is even now): \begin{equation}
     \begin{split}
         M_1&: \text{relabel},\\
         M_2&: \text{relabel}, \\
          M_3&: \text{relabel},\\
          \vdots \\
          M_{\frac{n}{2}}&: \text{relabel}, \\
           M_{\frac{n}{2}+1}&: \text{keep},\\
           M_{\frac{n}{2}+2}&: \text{keep}, \\
            \vdots \\
        M_{n}&: \text{keep},        
     \end{split}
 \end{equation} i.e., relabeling the outcomes of the first $\frac{n}{2}$ measurements.
 The behaviour we then obtain indeed satisfies Eq.~\eqref{eq:appendix_p_01_=0_forall_i} and Eq.~\eqref{eq:appendix_p_01__neqzero}.

\subsection{A logically contextual $n$-cycle behavior for even and odd $n$} 
We can thus conclude that we can write behaviors demonstrating $n$-cycle logical contextuality in a way such that the behavior for odd and even $n$ can be presented in a uniform way:
\begin{align} 
&p(0,1 \vert i,i+1)=0 \text{ for all $1 \leq i < n$}, \label{eq:appendix_for_all_n}
\end{align} and \begin{equation} \label{eq:appendix_uniform_nonzero}
    p(0,1 \vert 1, n) > 0. 
\end{equation}
If one assumes that the result of finding a box empty or full is predetermined and independent of which boxes are opened (i.e., noncontextuality), then a contradiction arises as follows. Let $m_i$ be the measurement outcome of opening box $i$. Eq.~\eqref{eq:appendix_uniform_nonzero} implies that we can find box 1 empty ($m_1=0$) and box $n$ full ($m_n=1$). Using $m_1 = 0$ in Eq.~\eqref{eq:appendix_for_all_n}, we find $m_2=0$. Using $m_2=0$ again in Eq.~\eqref{eq:appendix_for_all_n}, we find $m_3 = 0$. This goes on, with using $m_i = 0$ in Eq.~\eqref{eq:appendix_for_all_n} to conclude that $m_{i+1}=0$ for all $1 \leq i < n$, so that finally $m_n=0$, contradicting our initial post-selection on finding box $n$ full, $m_n=1$.

\section{Generalization to $n$-cycle extended Wigner's friend paradoxes} \label{sec:n_cycle_Wignerfriend}
Having described the 5-cycle Wigner's friend paradox, we now generalize to an $n$-cycle Wigner's friend argument for $n \geq 4$ a natural number. {Note that the correlations in the $n=4$ case are isomorphic to those appearing in the Hardy proof of Bell nonlocality, which is why we focused on the $n=5$ case here and in the main text.}  Note {also} that we use the labeling convention for measurement outcomes introduced in  Appendix~\ref{sec:n_cycle_logical_contextuality} and~\ref{appendix: from odd to even}. In particular, this labeling is distinct from that used in the main text.

The experimental set-up for this argument involves $n$ friends $A_1,A_2,\ldots,A_n$ and a superobserver Wigner. Consider a prepared system $S$ and measurements $M_1,\ldots,M_n$. Measurements $M_{i \mod n},M_{(i+1) \mod n}$ are compatible and thus can be jointly performed on $S$, and satisfy Eq.~\eqref{eq:cycle_for_all_n} and Eq.~\eqref{eq:cycle_n_neq0}. Correlations of this sort on a  system $S$ can be realized in quantum theory as shown in Ref.~\cite{santos2021conditions}. The $n$ friends $A_1,\ldots,A_n$ perform the measurements $M_1,\ldots,M_n$ sequentially on $S$, and at particular times the superobserver Wigner undoes the first $n-2$ of these measurements. By the Universality of Quantum Theory, agent $A_i$ performing measurement $M_i$ on $S$ can be modeled as a unitary $U_{M_i}$.

More specifically, first agent $A_1$ performs measurement $M_1$ on $S$, followed by agent $A_2$ performing measurement $M_2$ on $S$, after which superobserver Wigner undoes the first measurement by applying the unitary $U^\dagger_{M_1}$. Next, agent $A_3$ performs $M_3$, followed by Wigner undoing $A_2$'s measurement by applying $U^\dagger_{M_2}$. This continues, with agent $A_i$ performing measurement $M_i$ on $S$, followed by Wigner undoing $A_{i-1}$'s measurement by applying $U^\dagger_{M_{i-1}}$, until agent $A_n$ has performed their measurement. 

From Absoluteness of Observed Events, each agent $A_i$ observes an absolute outcome, denoted $a_i$, during their measurement $M_i$, with potential outcomes $0,1$.

\textit{No-go theorem.}---Measurements $M_1$ and $M_2$ are done immediately in sequence without any processes occurring in between, so the joint outcome $(a_1,a_2)$ can be directly observed (up until the time when $U_{M_2}^{\dagger}$ is performed). The Born rule predicts that
\begin{equation}
\label{eq_a12_n}
    p(a_1=0,a_2=1)=0,
\end{equation}
just as in \cref{eq:cycle_for_all_n}, the correlation for the $(1,2)$ pair of measurements in the $n$-cycle noncontextuality scenario. Moreover, this Born rule prediction---and those that follow---can be made without appealing to the projection postulate, as explained in the main text.

Measurements $M_2$ and $M_3$ are not done immediately in sequence; rather, the unitary $U_{M_1}^{\dagger}$ is applied in between them. Nevertheless, the joint outcome $(a_2,a_3)$ is still directly observable (up until $U_{M_2}^{\dagger}$ is performed). The Born rule predicts that
\begin{equation}
\label{eq_a23_n}
    p(a_2=0,a_3=1)=0,
\end{equation}
just as in \cref{eq:cycle_for_all_n}, the correlation for the $(2,3)$ pair of measurements in the $n$-cycle noncontextuality scenario.
This is the case because the unitary $U_{M_1}^{\dagger}$ commutes with the $M_3$ measurement, and so it does not affect the (observable) Born rule prediction for $p(a_2,a_3)$. 

Similarly, the Born rule predicts that 
\begin{align}
    p(a_i=0,a_{i+1}=1)=0 \text{ for all $1 \leq i < n$,} \label{eq_aii+1_n} 
\end{align}
since this refers only to joint outcomes that can be directly observed, and the only process happening between each pair of measurements is a unitary that commutes with those measurements (and so cannot affect their statistics). \cref{eq_aii+1_n} are analogous to \cref{eq:cycle_for_all_n} in the $n$-cycle noncontextuality scenario.

However, the same argument cannot be applied to the joint outcome $(a_1,a_n)$, since it necessarily cannot be observed by any observer, even in principle. This is because the outcome $a_1$ is erased (by the unitary $U_{M_1}^{\dagger}$) prior to the outcome $a_n$ being generated. Consequently, even though the process  $U:= U_{M_{n-2}}^{\dagger} U_{M_{n-1}} U_{M_{n-3}}^{\dagger} U_{M_{n-2}} \ldots U_{M_1}^{\dagger} U_{M_2}$, which is done between $M_1$ and $M_n$, commutes with $M_n$, the assertion that it does not affect the correlation between $M_1$ and $M_n$ is {\em not} guaranteed by the correctness of operational quantum theory (in particular, it does not follow from our assumption of Possibilistic Born Rule). To complete the argument, we use the assumption of Commutation Irrelevance, just as we did in the main text. Consequently,
assuming Commutation Irrelevance, the unobservable correlations for the actual experiment (where $U$ is done between $M_1$ and $M_n$) are  constrained in the same way:
\begin{equation}
    p(a_1=0,a_n=1) > 0. \label{eq_a1n}
\end{equation}
This is analogous to \cref{eq:cycle_n_neq0} in the $n$-cycle noncontextuality scenario.

While \cref{eq_aii+1_n} implies that the actually observed outcomes must satisfy
\begin{equation}
    a_1=0\Rightarrow a_2=0 \Rightarrow \ldots \Rightarrow a_n=0,
\end{equation} 
\cref{eq_a1n} tells us that the event $(a_1=0,a_n=1)$ occurs in some runs of the protocol.
Thus, for every $n\geq 4$ we have a contradiction and consequently a no-go theorem against the conjunction of our assumptions: Universality of Unitarity, Absoluteness of Observed Events, Possibilistic Born Rule and Commutation Irrelevance.

\section{Commutation between unitaries representing measurements}
\label{app_comUni}

Here, we show that the unitary representations of the measurements $M_i$ and $M_j$ commute for all of the pairs $(i,j)\in \{(1,2),(2,3),(3,4),(4,5),(1,5)\}$ and that, the unitary representations of the reversals of measurements $M_i$ and $M_j$ also commute for all of these pairs, as well as the unitary representations of $M_i$ and the reversal of $M_j$.

Specifically, we take the unitary representing $M_i$ to be the following CNOT gate
\begin{align}
   &U_{M_i}:={\rm CNOT_{SA_i}} \\ \nonumber
   =& (\Pi_{v_i})_{\rm S}\otimes\mathbb{1}_{\rm A_i}+ (\Pi_{v_i^{\perp}})_{\rm S}\otimes ( \ket{1}\bra{0} + \ket{0}\bra{1})_{\rm A_i},
\end{align}
where $\Pi_{v_i}:=\ket{v_i}\bra{v_i}$ and $\Pi_{v_i^{\perp}}:=\ket{v_i^{\perp}}\bra{v_i^{\perp}}$. As such, the reversal of the measurement, which is represented by $U_{M_i}^{\dagger}$ is the same as the gate CNOT$_{SA_i}$ defined above (Eq.~\eqref{eq_uni}).

Similarly, 
\begin{align}
   &U_{M_j}:={\rm CNOT_{SA_j}} \\ \nonumber
    =&(\Pi_{v_j})_{\rm S}\otimes\mathbb{1}_{\rm A_j}+ (\Pi_{v_j^{\perp}})_{\rm S}\otimes ( \ket{1}\bra{0} + \ket{0}\bra{1})_{\rm A_j}, 
\end{align}
where $\Pi_{v_j}:=\ket{v_j}\bra{v_j}$ and $\Pi_{v_j^{\perp}}:=\ket{v_j^{\perp}}\bra{v_j^{\perp}}$.

Then,
\begin{align}
    &(U_{M_i}\otimes\mathbb{1}_{A_j})(U_{M_j}\otimes\mathbb{1}_{A_i})\\ \nonumber
    =&(\Pi_{v_i}\Pi_{v_j})_{\rm S}\otimes\mathbb{1}_{A_i}\otimes\mathbb{1}_{A_j} \\ \nonumber
    + & (\Pi_{v_i}\Pi_{v_j^{\perp}})_{\rm S}\otimes\mathbb{1}_{A_i}\otimes ( \ket{1}\bra{0} + \ket{0}\bra{1})_{\rm A_j}  \\ \nonumber
    + & (\Pi_{v_i^{\perp}}\Pi_{v_j})_{\rm S}\otimes ( \ket{1}\bra{0} + \ket{0}\bra{1})_{\rm A_i} \otimes\mathbb{1}_{A_j} \\ \nonumber
    + & (\Pi_{v_i^{\perp}}\Pi_{v_j^{\perp}})_{\rm S}\otimes ( \ket{1}\bra{0} + \ket{0}\bra{1})_{\rm A_i} \otimes ( \ket{1}\bra{0} + \ket{0}\bra{1})_{\rm A_j}.
\end{align}
and
\begin{align}
    &(U_{M_j}\otimes\mathbb{1}_{A_i})(U_{M_i}\otimes\mathbb{1}_{A_j})\\ \nonumber
    =&(\Pi_{v_j}\Pi_{v_i})_{\rm S}\otimes\mathbb{1}_{A_i}\otimes\mathbb{1}_{A_j} \\ \nonumber
    + & (\Pi_{v_j^{\perp}}\Pi_{v_i})_{\rm S}\otimes\mathbb{1}_{A_i}\otimes ( \ket{1}\bra{0} + \ket{0}\bra{1})_{\rm A_j}  \\ \nonumber
    + & (\Pi_{v_j}\Pi_{v_i^{\perp}})_{\rm S}\otimes ( \ket{1}\bra{0} + \ket{0}\bra{1})_{\rm A_i} \otimes\mathbb{1}_{A_j} \\ \nonumber
    + &(\Pi_{v_j^{\perp}})_{\rm S}\otimes ( \ket{1}\bra{0} + \ket{0}\bra{1})_{\rm A_i} \otimes ( \ket{1}\bra{0} + \ket{0}\bra{1})_{\rm A_j}.
\end{align}

Since by definition,
\begin{align}
    [\Pi_{v_i},\Pi_{v_j}]=[\Pi_{v_i},\Pi_{v_j^{\perp}}]=[\Pi_{v_i^{\perp}},\Pi_{v_j}]=[\Pi_{v_i^{\perp}},\Pi_{v_j^{\perp}}],
\end{align}
we have 
\begin{equation}
    [U_{M_i}\otimes\mathbb{1}_{A_j},U_{M_j}\otimes\mathbb{1}_{A_i}]=0.
\end{equation}

\end{appendix}

\bibliography{refs}

\end{document}